\documentclass[aps,pre,twocolumn,superscriptaddress,showpacs]{revtex4-1}
\usepackage{graphicx}% Include figure files
\usepackage{dcolumn}% Align table columns on decimal point
\usepackage{bm}% bold math
\usepackage{epsfig}
\usepackage{color}
\begin{document}

%\preprint{}

\title{Studying the Kinetics of a Self-propelled Cruiser in 2D
  Granular Media under Gravity}

% repeat the \author .. \affiliation etc. as needed \email,
% \thanks, \homepage, \altaffiliation all apply to the current

\author{Guo-Jie J. Gao}
\email{koh.kokketsu@shizuoka.ac.jp, gjjgao@gmail.com}
%\thanks{}
\affiliation{Department of Mathematical and Systems Engineering,
  Shizuoka University, Hamamatsu, Shizuoka 432-8561, Japan}

\date{\today}

\begin{abstract}
We propose a cruiser able to move in a granular medium made of nearly
50-50 bidisperse dissipative particles under gravity. The cruiser has
a circular shape with a square indentation on its edge. By shifting
and then ejecting granular particles entering its indent-region facing
a given direction, the cruiser gains thrust to push itself forward in
the same direction, which can be either perpendicular or parallel to
gravity. Using molecular dynamics (MD) simulations, we identify three
universal phases during one particle-ejection process: 1) acceleration
by the ejection thrust, 2) deceleration by the compressed particles
ahead and 3) relaxation with the decompressed particles. We also
confirm that the cruising capability improves with increasing the
particle-ejection strength and with decreasing the interference from
gravity.
\end{abstract}

%\pacs{61.43.-j,61.46.-w,81.05.-t}
%\keywords{}

\maketitle

\section{Introduction}
\label{introduction}
Designing and analyzing self-propelled machines in thermal fluids such
as air and water can be dated back to the beginning of recorded
history. The equilibrium characteristics of thermal fluids allow us to
easily understand their physical properties and explore their
interactions with objects inside them both microscopically and
macroscopically. However, similar endeavors to study mobile machines
in athermal fluids such as nonequilibrium granular materials have been
initiated only recently \cite{goldman16}, mainly because the
nonequilibrium features make developing a governing equation of the
system a challenging task \cite{behringer96}. Among the finite
attempts tackling this challenge in the literature, except some
artificial designs \cite{goldman11, goldman15_2, kamrin16, kamrin17,
  pak17, gao18}, most of them are bio-inspired, such as mimicking
sandfish, lizard and snake\cite{herrmann09, goldman09_2, goldman13,
  goldman14, goldman15_1, goldman15_2, pak16, goldman18}, insects
\cite{goldman09_1} and clam that uses an intricate way to move forward
by swallowing and discharging sands \cite{hosoi14,
  goldman15}. Learning from the nature has the benefit of always
having an original counterpart that has been optimised through long
evolution to compare with, and the mimicked designs are guaranteed to
function under known conditions. Nevertheless, the bio-inspired
designs have their limitation. Sometime to achieve better motion
efficiency or controllability, one needs to introduce mechanical
components that do not exist in living creatures such as installing a
propeller or a compressor on a glider imitating a bird initially.

In this study, we propose an extremely simple and novel design of a
self-propelled cruiser, able to move freely in a 2D granular medium
made of bidisperse dissipative particles under gravity. The cruiser
has a circular shape plus a square indentation of finite size on its
edge. To move into a given direction, the cruiser directs its
indentation facing the moving direction, shifts any particles entering
the indentation to its rear-half part and ejects the shifted particles
backward to gain forward thrust. Through successive ejections, the
cruiser can propel itself and travel within the granular medium either
perpendicular or parallel to gravity. Using molecular dynamics (MD)
methods, we test the kinetic response of the cruiser under different
ejection strengths and study its maneuverability within frictionless
bidisperse particles with damping. We identify three distinct and
universal phases during one particle-ejection: 1) obtaining momentum
from the ejected particles, 2) proceeding due to the inertial effects
and 3) relaxing with the rebounding particles compressed
ahead. Moreover, we also confirm that the cruiser can move further
proportional to the ejection strength. Finally, cruising against or
with gravity reduces the propelling efficiency.

Below we elaborate on the details of the simulated system and the
design of our self-propelled cruiser in section \ref{method}, followed
by quantitative analysis of its ejection kinetics and cruising
capability under different simulation setups in section
\ref{results_and_discussions}. We conclude our study in section
\ref{conclusions}.

\section{Numerical simulation method}
\label{method}

\subsection{Preparing a non-overlapped initial configuration}
\label{System geometry}
\begin{figure}
\includegraphics[width=0.40\textwidth]{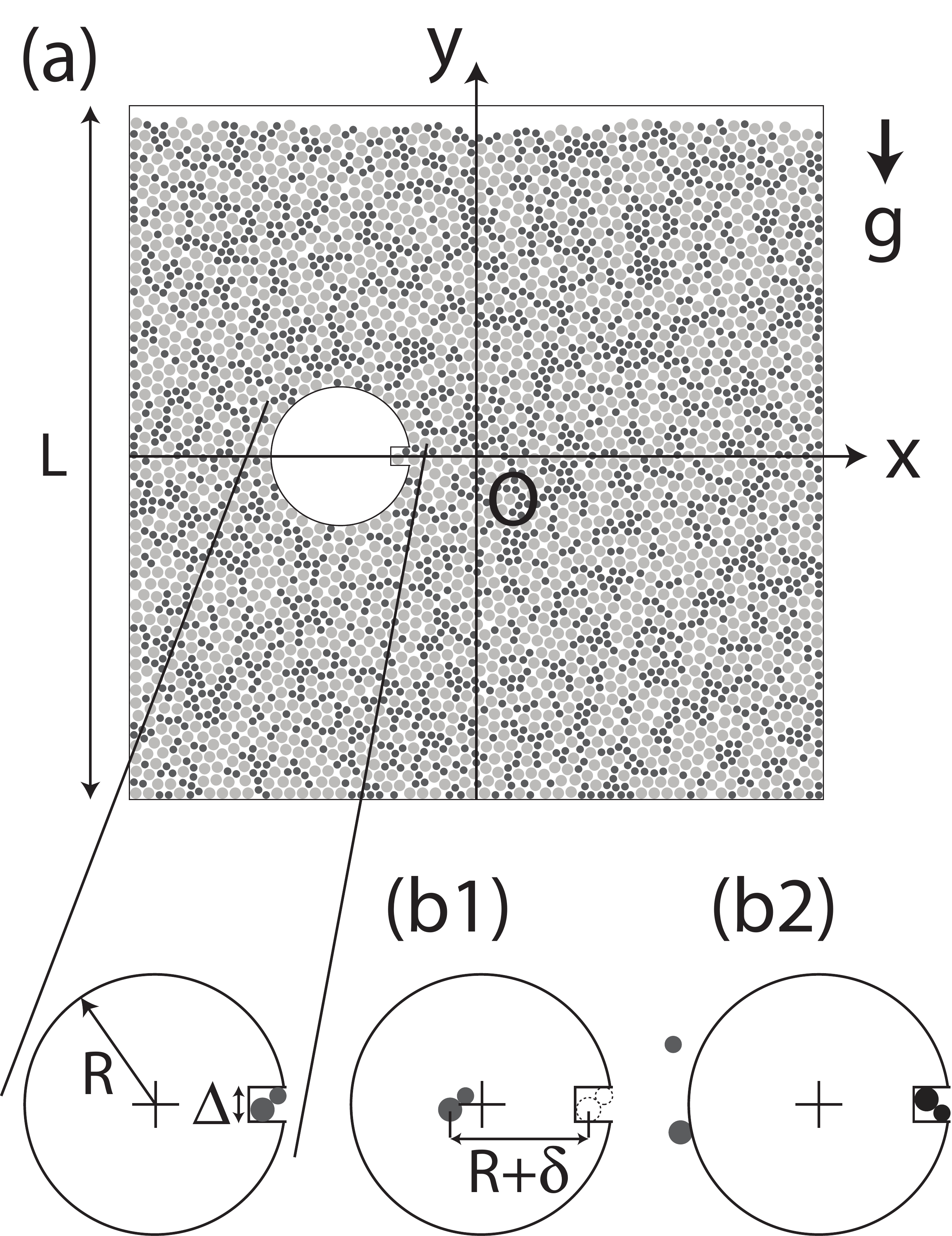}
\caption{\label{fig:scheme} (Color online) (a) The MD simulation setup
  of a square container of size $L$, storing about $4000$ bidisperse
  dissipative particles and a circular cruiser under gravity $g$. The
  small (dark grey) and large (light grey) particles measure $d_s$ and
  $d_l$ in diameter, separately, with a size ratio of $1.4$. The
  blowup shows the cruiser, having a radius $R$ and an square
  indentation of size $\Delta=2d_s$ on its edge. The cruiser shifts
  particles entering its square indent-region to its rear by a
  distance of $R+\delta$ and ejects them backward to gain thrust for
  moving forward, as shown in (b1) and (b2), respectively.}
\end{figure}

Using molecular dynamics (MD) method, we study the kinetics of a
self-propelled cruiser moving in a granular medium of nearly 50-50
bidisperse frictionless circular dry particles with damping in a
square container of size $L$, as shown in Fig. \ref{fig:scheme}(a).
 
To prepare a non-overlapped initial configuration for the MD
simulation, in the square container initially we randomly place $N_s$
small discs and $N_l$ large discs of diameters $d_s$ and $d_l$,
respectively, determined by $\phi = \left( {{\raise0.7ex\hbox{$\pi $}
    \!\mathord{\left/ {\vphantom {\pi
          {4{L^2}}}}\right.\kern-\nulldelimiterspace}
    \!\lower0.7ex\hbox{${4{L^2}}$}}} \right)({N_s}{d_s}^2 +
{N_l}{d_l}^2) = 0.833$, and $N_{tot}=N_s+N_l=4096$. This gives ${d_s}
\approx 0.0132L$. We keep the diameter ratio $d_l/d_s$ between large
and small discs at $1.4$ to prevent artificial crystallization in a
two dimensional environment.

Then we insert a circular cruiser of radius $R=0.1L$ with a
finite-sized square indentation of size $\Delta=2d_s$ on its edge into
the container and remove all granular particles within a circular
range $R$ from the center of the cruiser. On average, this removes
about $100 \sim 200$ particles from the container and evenly reduces
$N_s$ and $N_l$, which produces a nearly 50-50 bidisperse system, and
$N_{tot}$ becomes slightly less than $4000$.

Finally, to quickly remove all inter-particle overlaps in the system,
we perform MD simulations under zero gravity by introducing a
finite-range, pairwise purely repulsive linear spring force
$\mathord{\buildrel{\lower3pt\hbox{$\scriptscriptstyle\rightharpoonup$}}
  \over f} _{ij}^{n}({r_{ij}})$ and minimizing the total normal force
on each object in the system, where the normal interaction between any
two objects $i$ and $j$ (particle-particle or particle-cruiser) is
governed by
\begin{equation} \label{object_force}
\mathord{\buildrel{\lower3pt\hbox{$\scriptscriptstyle\rightharpoonup$}}
  \over f} _{ij}^n({r_{ij}}) = \frac{\epsilon}{{d_{ij}^2}}{\delta
  _{ij}}\Theta ({\delta _{ij}}){{\hat r}_{ij}},
\end{equation}
where $r_{ij}$ is the separation between objects $i$ and $j$,
$\epsilon$ is the characteristic elastic energy scale, $d_{ij} =
(d_i+d_j)/2$ is the average size of objects $i$ and $j$,
$\delta_{ij}=d_{ij}-r_{ij}$ is the inter-object overlap, $\Theta(x)$
is the Heaviside step function, and ${{\hat r}_{ij}}$ is the unit
vector connecting the centers of the two objects. When approaching a
wall, a particle feels its identical image with the same relative
distance on the other size of the wall.

%%%%%%%%%%%%%%%%%%%%%%
%%%%%%%%%%%%%%%%%%%%%%

\subsection{System setup for cruising under gravity}
\label{Interactions}

\subsubsection{Forces on a granular particle}
\label{part_forces}
In the simulation under nonzero gravity, each frictionless particle
$i$ obeys Newton's translational equation of motion
\begin{equation} \label{newton_law}
{{\mathord{\buildrel{\lower3pt\hbox{$\scriptscriptstyle\rightharpoonup$}}
      \over F} }_i} =
\mathord{\buildrel{\lower3pt\hbox{$\scriptscriptstyle\rightharpoonup$}}
  \over F} _i^{{\mathop{\rm int}} } +
\mathord{\buildrel{\lower3pt\hbox{$\scriptscriptstyle\rightharpoonup$}}
  \over F} _i^W +
\mathord{\buildrel{\lower3pt\hbox{$\scriptscriptstyle\rightharpoonup$}}
  \over F} _i^C +
\mathord{\buildrel{\lower3pt\hbox{$\scriptscriptstyle\rightharpoonup$}}
  \over F} _i^G =
        {m_i}{{\mathord{\buildrel{\lower3pt\hbox{$\scriptscriptstyle\rightharpoonup$}}
              \over a} }_i},
\end{equation}
where
${{\mathord{\buildrel{\lower3pt\hbox{$\scriptscriptstyle\rightharpoonup$}}
      \over F} }_i}$ is the total force acting on particle $i$ with
mass $m_i$ and acceleration
${{\mathord{\buildrel{\lower3pt\hbox{$\scriptscriptstyle\rightharpoonup$}}
      \over a}
  }_i}$. $\mathord{\buildrel{\lower3pt\hbox{$\scriptscriptstyle\rightharpoonup$}}
  \over F} _i^{{\mathop{\rm int}} }$,
$\mathord{\buildrel{\lower3pt\hbox{$\scriptscriptstyle\rightharpoonup$}}
  \over F} _i^W$,
$\mathord{\buildrel{\lower3pt\hbox{$\scriptscriptstyle\rightharpoonup$}}
  \over F} _i^C$ and
$\mathord{\buildrel{\lower3pt\hbox{$\scriptscriptstyle\rightharpoonup$}}
  \over F} _i^G$ are forces acting on the particle from its contact
neighbors, the container walls, the cruiser and gravity,
respectively. Below we elaborate on each of these terms individually.

To simulate frictionless granular materials, we consider only the
interparticle normal forces for simplicity reason \cite{gao09}. The
interparticle force
$\mathord{\buildrel{\lower3pt\hbox{$\scriptscriptstyle\rightharpoonup$}}
  \over F} _i^{{\mathop{\rm int}} }$ on particle $i$ having $N_c$
contact neighbors $j$ can be expressed as
\begin{equation} \label{interparticle_force_law}
\mathord{\buildrel{\lower3pt\hbox{$\scriptscriptstyle\rightharpoonup$}}
  \over F} _i^{{\mathop{\rm int}} } = \sum\limits_{j \ne i}^{{N_c}}
        {[\mathord{\buildrel{\lower3pt\hbox{$\scriptscriptstyle\rightharpoonup$}}
              \over f} _{ij}^n} ({r_{ij}}) +
          \mathord{\buildrel{\lower3pt\hbox{$\scriptscriptstyle\rightharpoonup$}}
            \over f} _{ij}^{d}({r_{ij}})],
\end{equation}
where
$\mathord{\buildrel{\lower3pt\hbox{$\scriptscriptstyle\rightharpoonup$}}
  \over f} _{ij}^n({r_{ij}})$ is the interparticle normal force,
having the same form defined in Eqn.(\ref{object_force}), and
$\mathord{\buildrel{\lower3pt\hbox{$\scriptscriptstyle\rightharpoonup$}}
  \over f} _{ij}^d({r_{ij}})$ is the interparticle normal damping
force defined in Eqn.(\ref{particle_damping_force}) below. We consider
the interparticle normal damping force proportional to the relative
velocity between particles $i$ and $j$
\begin{equation} \label{particle_damping_force}
\mathord{\buildrel{\lower3pt\hbox{$\scriptscriptstyle\rightharpoonup$}}
  \over f} _{ij}^d({r_{ij}}) = - b\Theta ({\delta
  _{ij}})({{\mathord{\buildrel{\lower3pt\hbox{$\scriptscriptstyle\rightharpoonup$}}
      \over v} }_{ij}} \cdot {{\hat r}_{ij}}){{\hat r}_{ij}},
\end{equation}
where $b$ is the damping parameter,
${{\mathord{\buildrel{\lower3pt\hbox{$\scriptscriptstyle\rightharpoonup$}}
      \over v} }_{ij}}$ is the relative velocity between the two
particles. The normal damping force results in deduction of the
kinetic energy of the involved particles after each pairwise
collision.

The interaction force
$\mathord{\buildrel{\lower3pt\hbox{$\scriptscriptstyle\rightharpoonup$}}
  \over F} _i^W$ between particle $i$ and a container wall has an
analogous form to the interparticle normal interaction
$\mathord{\buildrel{\lower3pt\hbox{$\scriptscriptstyle\rightharpoonup$}}
  \over f} _{ij}^{{\mathop{\rm n}} }$, except with
$\epsilon^W=2\epsilon$, which means when a particle hits a wall, it
experiences a repulsive force as if it hit another mirrored self on
the other side of the
wall. $\mathord{\buildrel{\lower3pt\hbox{$\scriptscriptstyle\rightharpoonup$}}
  \over F} _i^W$ can include contributions from multiple walls: for
example, a particle can sit at the corner formed by two perpendicular
walls.

The particle-cruiser interaction force
$\mathord{\buildrel{\lower3pt\hbox{$\scriptscriptstyle\rightharpoonup$}}
  \over F} _i^C$ can be expressed as
\begin{equation} \label{part_cruiser_force}
\mathord{\buildrel{\lower3pt\hbox{$\scriptscriptstyle\rightharpoonup$}}
  \over F} _i^C =
\mathord{\buildrel{\lower3pt\hbox{$\scriptscriptstyle\rightharpoonup$}}
  \over f} _{iC}^n +
\mathord{\buildrel{\lower3pt\hbox{$\scriptscriptstyle\rightharpoonup$}}
  \over f} _{iC}^d,
\end{equation}
where
$\mathord{\buildrel{\lower3pt\hbox{$\scriptscriptstyle\rightharpoonup$}}
  \over f} _{iC}^n$ and
$\mathord{\buildrel{\lower3pt\hbox{$\scriptscriptstyle\rightharpoonup$}}
  \over f} _{iC}^d$ are the particle-cruiser normal force and normal
damping force, respectively. If a particle touches the circular edge
of the cruiser,
$\mathord{\buildrel{\lower3pt\hbox{$\scriptscriptstyle\rightharpoonup$}}
  \over f} _{iC}^n =
\mathord{\buildrel{\lower3pt\hbox{$\scriptscriptstyle\rightharpoonup$}}
  \over f} _{ij}^n$, and
$\mathord{\buildrel{\lower3pt\hbox{$\scriptscriptstyle\rightharpoonup$}}
  \over f} _{iC}^d$ is zero. Otherwise, if a particle touches the
walls of the square indentation of the cruiser,
$\mathord{\buildrel{\lower3pt\hbox{$\scriptscriptstyle\rightharpoonup$}}
  \over f} _{iC}^n =
\mathord{\buildrel{\lower3pt\hbox{$\scriptscriptstyle\rightharpoonup$}}
  \over F} _i^W$ and
$\mathord{\buildrel{\lower3pt\hbox{$\scriptscriptstyle\rightharpoonup$}}
  \over f} _{iC}^d =
\mathord{\buildrel{\lower3pt\hbox{$\scriptscriptstyle\rightharpoonup$}}
  \over f} _{ij}^d$, except with a much larger damping parameter
$b_C=50b$. The large $b_C$ helps the cruiser to keep granular
particles within its square indent-region once they enter it.

Finally,
$\mathord{\buildrel{\lower3pt\hbox{$\scriptscriptstyle\rightharpoonup$}}
  \over F} _i^G = - {m_i}g\hat y$, where $g$ is the gravitational
constant, and $\hat y$ is the unit vector in the upward
direction. There is no tangential interaction on frictionless
particles in this model, and therefore Newton's rotational equation of
motion is automatically satisfied.

\subsubsection{Forces on the cruiser}
\label{cruiser_forces}
Based on Newton's translational equation of motion, the total force
${{\mathord{\buildrel{\lower3pt\hbox{$\scriptscriptstyle\rightharpoonup$}}
      \over F} }_C}$ acting on the cruiser with mass $m_C$ and
acceleration
${{\mathord{\buildrel{\lower3pt\hbox{$\scriptscriptstyle\rightharpoonup$}}
      \over a} }_C}$, in contact with $N_p$ particles, is
\begin{equation} \label{cruiser_tot_force}
{{\mathord{\buildrel{\lower3pt\hbox{$\scriptscriptstyle\rightharpoonup$}}
      \over F} }_C} = (-\sum\limits_i^{{N_p}} {
  \mathord{\buildrel{\lower3pt\hbox{$\scriptscriptstyle\rightharpoonup$}}
    \over F} _i^C)} +
\mathord{\buildrel{\lower3pt\hbox{$\scriptscriptstyle\rightharpoonup$}}
  \over F} _C^G =
        {m_C}{{\mathord{\buildrel{\lower3pt\hbox{$\scriptscriptstyle\rightharpoonup$}}
              \over a} }_C},
\end{equation}
where the minus sign in front of
${\mathord{\buildrel{\lower3pt\hbox{$\scriptscriptstyle\rightharpoonup$}}
    \over F} _i^C}$ comes from Newton's third law of motion, and
$\mathord{\buildrel{\lower3pt\hbox{$\scriptscriptstyle\rightharpoonup$}}
  \over F} _C^G = - {m_C}g\hat y$. In this study, we do not perform
simulations where the cruiser can touch the container wall, so the
cruiser-wall interaction can be safely ignored. 

We also do not consider Newton's rotational equation of motion of the
cruiser by redefining the indent-region of the cruiser according to
its center position at each MD step. This approach causes the cruiser
to drift slightly if any particle interacts with the indentation walls
of the cruiser. The drifting effect is not significant, because at
most only two or three particles can enter the indent-region based on
our MD simulation results. Besides, in the MD simulations, the mass of
an object is proportional to its area size, meaning the cruiser is
about $229$ or $117$ times heavier than a small or large granular
particle. The above two reasons justify this approach, so we can focus
only on the translational motion of the cruiser.

\subsubsection{Reference scales and simulation parameters}
\label{scales_parameters}
The MD simulations in this study use the diameter $d_s$ and mass $m_s$
of the small particles and the interparticle elastic potential
amplitude $\epsilon$ as the reference length, mass, and energy scales,
respectively. We choose the dimensionless damping parameter
$b^*=db/\sqrt{m\epsilon}=0.5$, the dimensionless gravity
$g^*=10^{-3}$, and a dimensionless time step $dt^*=dt/{d}\sqrt
{{m}/\epsilon}=10^{-2}$ throughout this study.

\subsubsection{Relaxation of the non-overlapped initial configuration}
\label{relax_IC}
After generating a non-overlapped initial configuration, we relax it
under nonzero gravity. To do this, we monitor the relative error
$\eta$ of the total potential energy of the system
$V_{tot}=V^{int}+V^W+V^G$ between time $t$ and $t+\Delta t$, where
$V^{int}$ is the total repulsive linear spring potential between
objects, $V^W$ is the total repulsive linear spring potential between
particles and walls, $V^G$ is the total gravitational potential in the
system, based on the forces given in section \ref{part_forces} and
section \ref{cruiser_forces}, and $\Delta t=1000$. We terminate the
relaxation when $\eta(V_{tot}) < 10^{-5}$. A relaxed initial
configuration and the cruiser with a radius $R$ and a square
indentation of size $\Delta=2d_s$, are shown in
Fig. \ref{fig:scheme}(a).

\subsection{Ejection mechanism of the cruiser}
\label{mechanism}
The ejection mechanism of the cruiser can be divided into two steps:
1) The cruiser detects if there are particles whose centers are within
the pocket region of its square indentation. If the answer is yes,
those particles will be shifted horizontally or vertically by a
distance $R+\delta$ away from the indentation, as depicted in
Fig. \ref{fig:scheme}(b1). $\delta$ is a control parameter in this
study. This shift of pocketed particles has twofold important purpose
in the coming relaxation step: First, we store spring potential energy
between the shifted particles and the cruiser, which will transform
into work and supply thrust to push the cruiser forward. Second, the
emptied indent-region offers necessary space to unjam the region ahead
of the cruiser a bit, so that the cruiser can proceed. 2) Using the
stored potential energy, the cruiser ejects the shifted particles
backward and obtain thrust to propel itself forward. The thrust has
the same form as
$\mathord{\buildrel{\lower3pt\hbox{$\scriptscriptstyle\rightharpoonup$}}
  \over f} _{ij}^n$, defined in Eqn.(\ref{object_force}). We do not
strictly control the ejection directions of the shifted particles and
let
$\mathord{\buildrel{\lower3pt\hbox{$\scriptscriptstyle\rightharpoonup$}}
  \over f} _{ij}^n$ be in charge of the process automatically. The
second step is done by a full relaxation of the system, terminated at
$\eta(V_{tot}) < 10^{-5}$. During the relaxation, New particles can
enter the pocket region of the indentation and stay there after the
relaxation ends, as shown in Fig. \ref{fig:scheme}(b2). The ejected
particles also prevent the cruiser from receding. By repeating the
above two steps, the cruiser can transit in the bidisperse granular
medium under systematic control.

\section{Results and Discussions}
\label{results_and_discussions}
Below we investigate the kinetics of three different particle-ejection
modes of the cruiser: 1) perpendicular to gravity, 2) against gravity
and 3) with gravity, followed by an analysis of the averaged
performance of the cruiser if it ejects pocketed-particles
successively in each mode at different ejection strength.

\subsection{Particle-ejection perpendicular to gravity}
\label{force_modeR}
\begin{figure}
\includegraphics[width=0.40\textwidth]{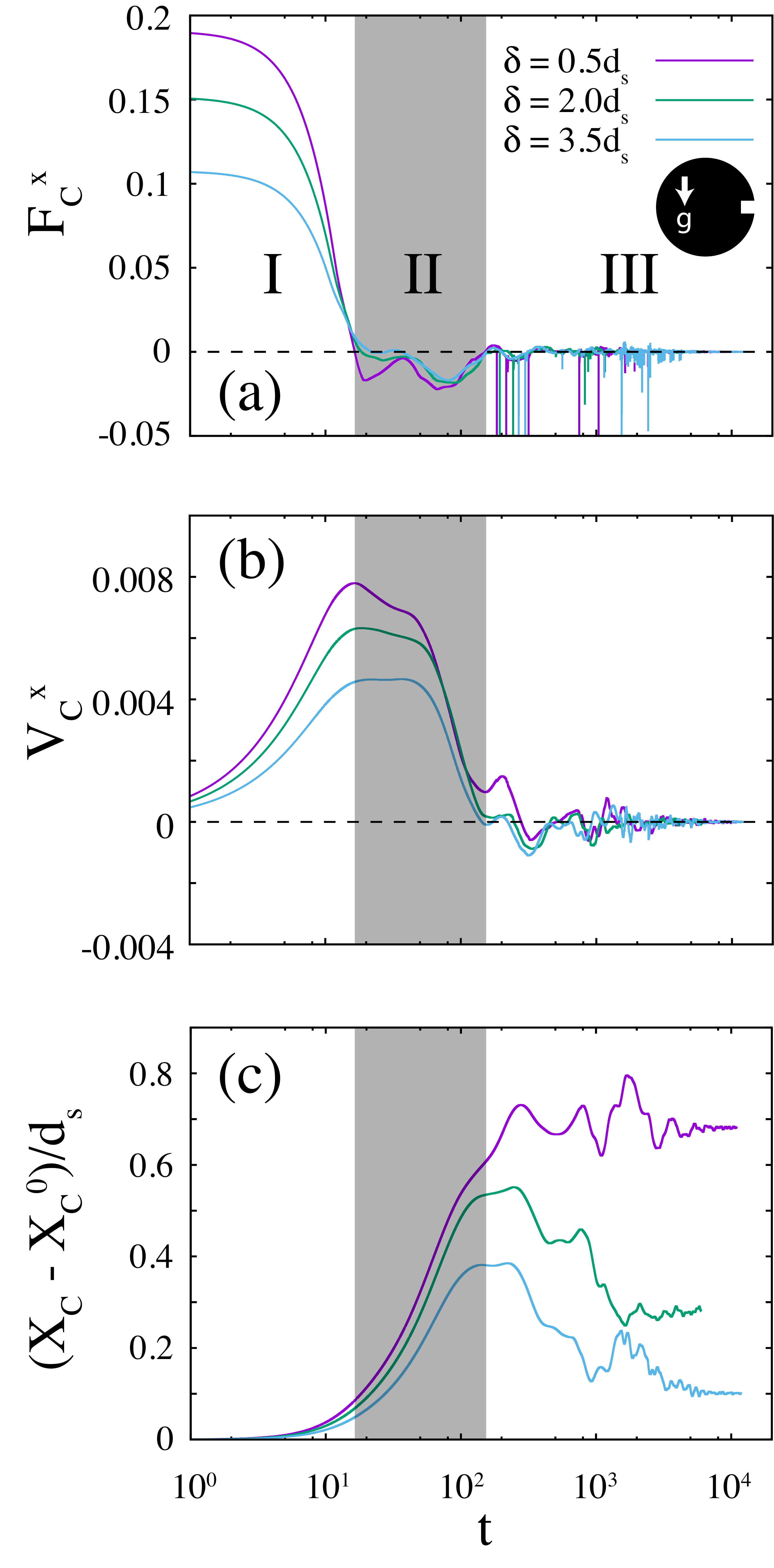}
\caption{\label{fig:modeR_details} (Color online) The $x$ components
  of (a) total force $F_C^x$ on the cruiser, (b) its velocity $V_C^x$
  and (c) its net displacement $(X_C-X_C^0)/d_s$, normalized by the
  small particle diameter $d_s$, during a single ejection of exactly
  two particles, perpendicular to gravity, as a function of the shift
  distance $\delta=0.5d_s$ (dark), $2.0d_s$ (medium) and $3.5d_s$
  (light), respectively. The particle-ejection strength is inversely
  proportional to $\delta$. The cruiser is initially placed at
  $(x,y)=(-0.2L,0.0)$ with its indentation pointing horizontally. The
  ejection process can be divided into three phases, I, II and III, as
  discussed in the text.}
\end{figure}

To study the kinetic response of the horizontal particle-ejection mode
perpendicular to gravity, initially we place the cruiser at
$(x,y)=(-0.2L,0.0)$ in the container and put the square indentation
facing the moving direction along the horizontal axis. After the first
relaxation, several particles enter the square indent-region of the
cruiser, and then we shift them horizontally by a distance of
$R+\delta$ away from the indentation, as depicted in
Fig. \ref{fig:scheme}(b1). We vary the value of $\delta$ from
$0.5d_s$, $2.0d_s$ to $3.5d_s$ and measure the total force $F_C^x$ on
the cruiser, its velocity component perpendicular to gravity $V_C^x$
and its normalized horizontal displacement $(X_C-X_C^0)/d_s$ when the
cruiser ejects the shifted particles in the second relaxation. In
Fig. \ref{fig:modeR_details}, we show these quantities from an
exemplary case as a function of time $t$, where exactly two particles
are ejected by the cruiser. We can separate the time series into three
phases, I, II, and III as follows.

%%%%%%%%%%%%%%%%%%%%%%
%%%%%%%%%%%%%%%%%%%%%%

Phase I in Fig. \ref{fig:modeR_details}(a) shows that initially
$F_C^x$ decreases monotonically and smoothly from a high value, when
the shifted particles are just being ejected from the cruiser, to a
nearly zero value, when they completely disengage from it. In this
phase, the cruiser picks up speed, and $V_C^x$ increases and
eventually reaches its maximum, as shown in
Fig. \ref{fig:modeR_details}(b). $V_C^x$ cannot increase anymore after
phase I, because the ejected particles have no more overlap with the
cruiser and therefore can supply no more thrust. The corresponding
$(X_C-X_C^0)/d_s$, however, only increases slightly, as shown in
Fig. \ref{fig:modeR_details}(c). This also justifies the nearly zero
$F_C^x$ mentioned above.

Next, in phase II, the cruiser keeps moving forward due to inertial
effects, using the momentum built up in the previous phase. The
forward-going cruiser keeps pressing the particles ahead, which cause
$F_C^x$ on the cruiser to become negative and against its motion, and
eventually the negative $F_C^x$ stops the cruiser. Nevertheless, it is
in this phase when new particles, used for next particle-ejection,
enter the indent-region, and the cruiser gains most of its forward
displacement among all three phases.

Finally, in phase III, the cruiser already exhausts most of its
momentum and the compressed particles ahead of it start to push
back. We can observe that the compressed particles decompress
themselves through a series of intermittent negative impulses. Each of
these impulses is not strong enough to push the cruiser back
substantially, and the cruiser never regain considerable momentum as
in phase I. The occasionally increased kinetic energy in the system
due to an impulse is quickly dissipated through interparticle damping
forces. Therefore, $V_C^x$ only fluctuates around zero. Until the end
of the relaxation, the accumulated backward movement is still not
large enough to send the cruiser back to where it started. The net
effect on $(X_C-X_C^0)/d_s$ is that the cruises can proceed by this
'one steps forward, half step back' strategy. As we reduce the
ejection strength $\delta$ from $0.5d_s$, $2.0d_s$ to $3.5d_s$, the
responses of all three quantities diminish systematically, but all
trends discussed above stay unchanged.

\subsection{Particle-ejection against or with gravity}
\label{force_modeU_modeD}
%%%%%%%%%%%%%%%%%%%%%%
\begin{figure}
\includegraphics[width=0.40\textwidth]{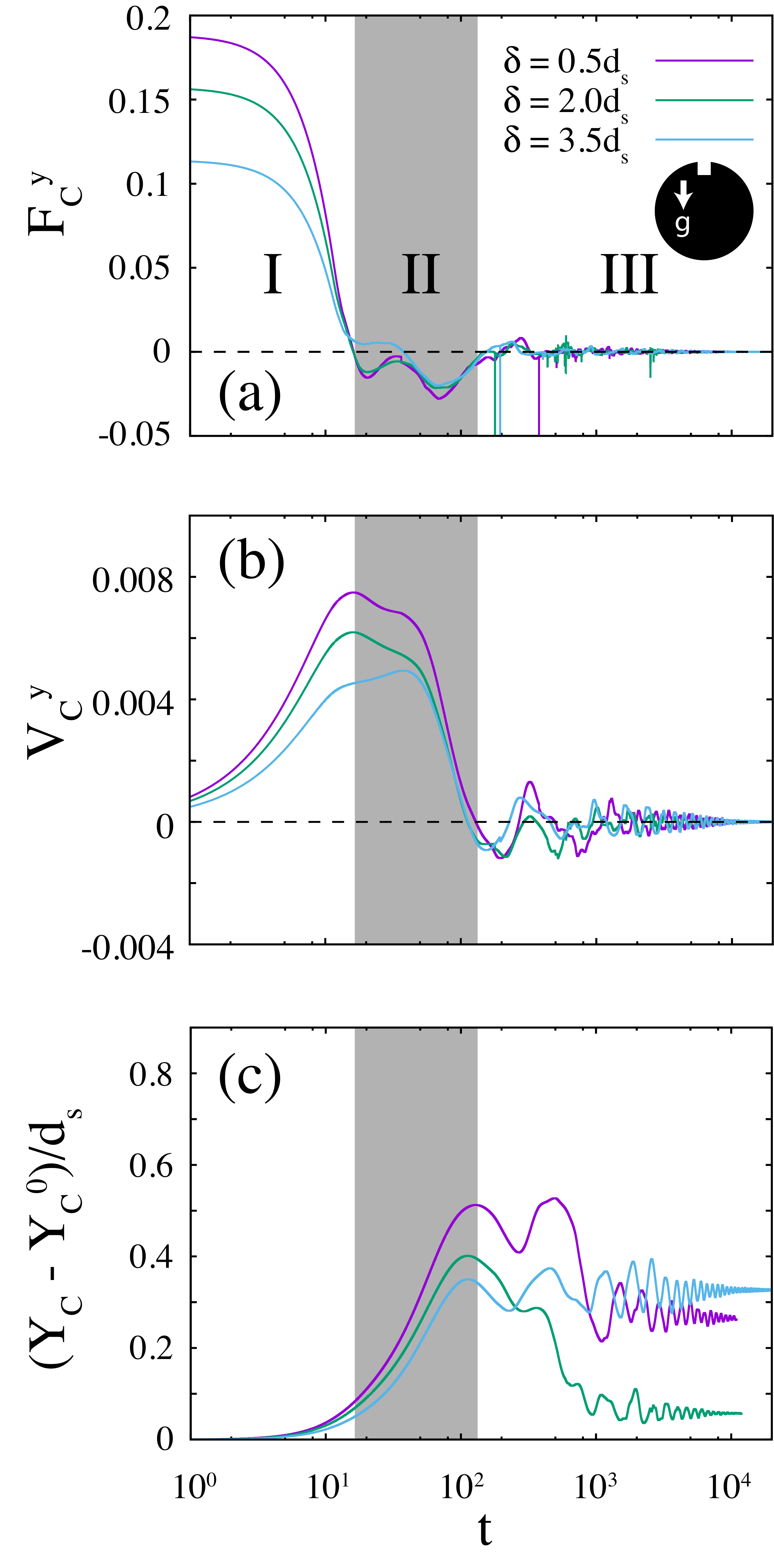}
\caption{\label{fig:modeU_details} (Color online) The $y$ components
  of (a) total force $F_C^y$ on the cruiser, (b) its velocity $V_C^y$
  and (c) its net displacement $(Y_C-Y_C^0)/d_s$, normalized by the
  small particle diameter $d_s$, during a single ejection of exactly
  two particles against gravity, as a function of the shift distance
  $\delta=0.5d_s$ (dark), $2.0d_s$ (medium) and $3.5d_s$ (light),
  respectively. The cruiser is initially placed at $(x,y)=(0.0,-0.2L)$
  with its indentation placed on the top of the cruiser. The ejection
  process can also be divided into three phases, I, II and III, as in
  Fig. \ref{fig:modeR_details}.}
\end{figure}
%%%%%%%%%%%%%%%%%%%%%%

%%%%%%%%%%%%%%%%%%%%%%
\begin{figure}
\includegraphics[width=0.40\textwidth]{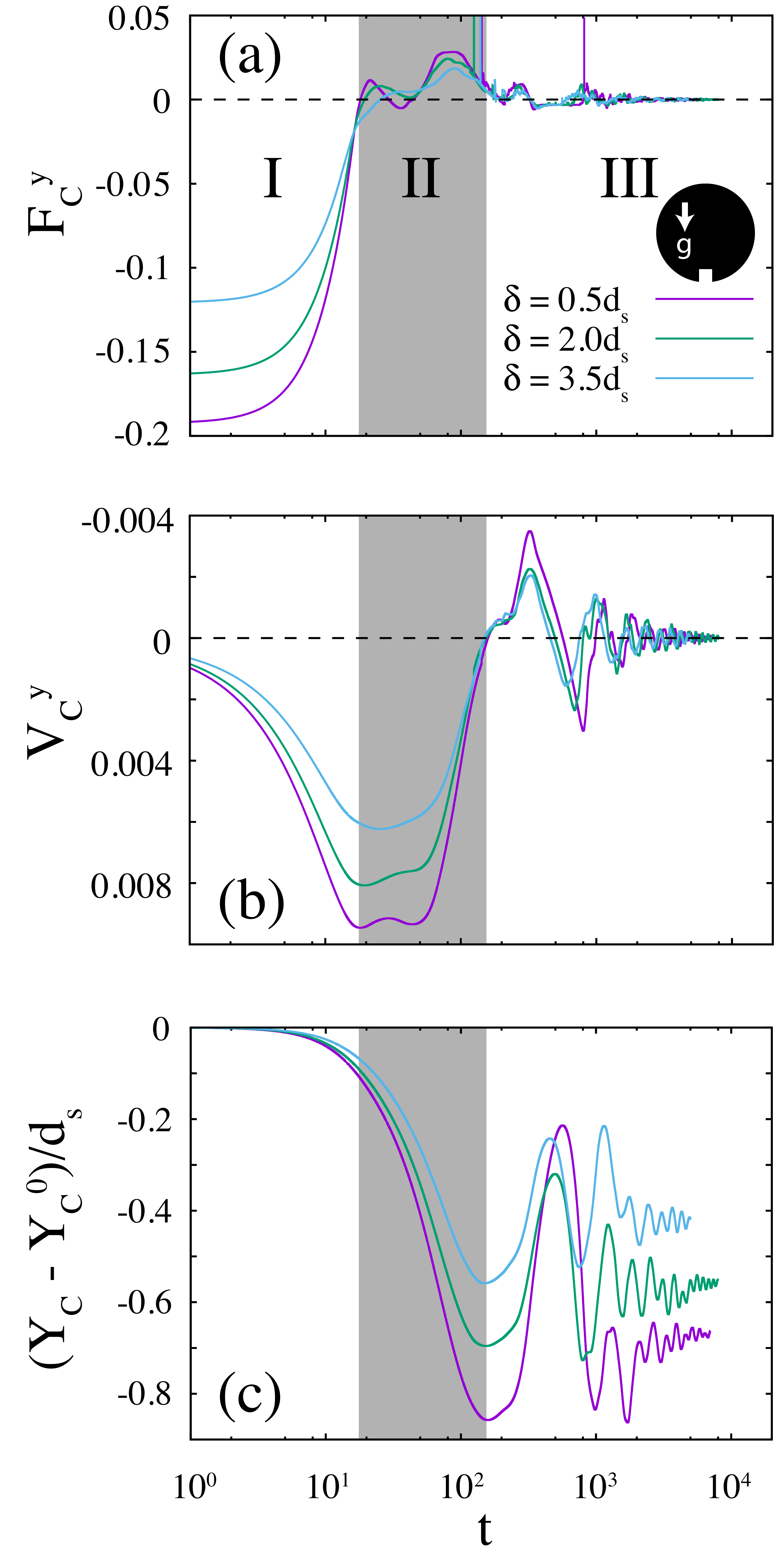}
\caption{\label{fig:modeD_details} (Color online) Same quantities as
  in Fig. \ref{fig:modeU_details} during a single ejection with
  gravity. The cruiser is initially placed at $(x,y)=(0.0,0.2L)$ with
  its indentation placed at the bottom of the cruiser.}
\end{figure}
%%%%%%%%%%%%%%%%%%%%%%

Similarly, to study the kinetics of the vertical particle-ejection
mode against or with gravity, initially we place the cruiser at
$(x,y)=(0.0,-0.2L)$ or $(x,y)=(0.0,0.2L)$ in the container and put the
square indentation on top of or at the base of the cruiser along the
vertical axis. Likewise, the behavior of $F_C^y$, $V_C^y$ and
$(Y_C-Y_C^0)/d_s$ as a function of time $t$ from exemplary cases,
where exactly two particles are ejected by the cruiser, are shown in
Fig. \ref{fig:modeU_details} and Fig. \ref{fig:modeD_details} for
tests against gravity and with gravity, respectively.

Basically, for each ejection mode, we observe similar patterns in all
three phases as those shown in the horizontal ejection mode, except
two new things to notice. First, in phase II, when the ejection mode
is in line with gravity, gravity also contributes to accelerate the
cruiser, and therefore the cruiser gains higher maximum velocity
$\left| {V_C^y} \right|$ than ejection against gravity, as can be seen
in Fig. \ref{fig:modeU_details}(b) and
Fig. \ref{fig:modeD_details}(b). Second, as expected, gravity makes
$(Y_C-Y_C^0)/d_s$ oscillate more obviously in phase III in both
vertical ejection modes, as shown in Fig. \ref{fig:modeU_details}(c)
and Fig. \ref{fig:modeD_details}(c), which is less noticeable in the
horizontal ejection mode.

\subsection{Cruising capability by successive particle-ejection}
\label{cruising_capability}
\begin{figure}
\includegraphics[width=0.40\textwidth]{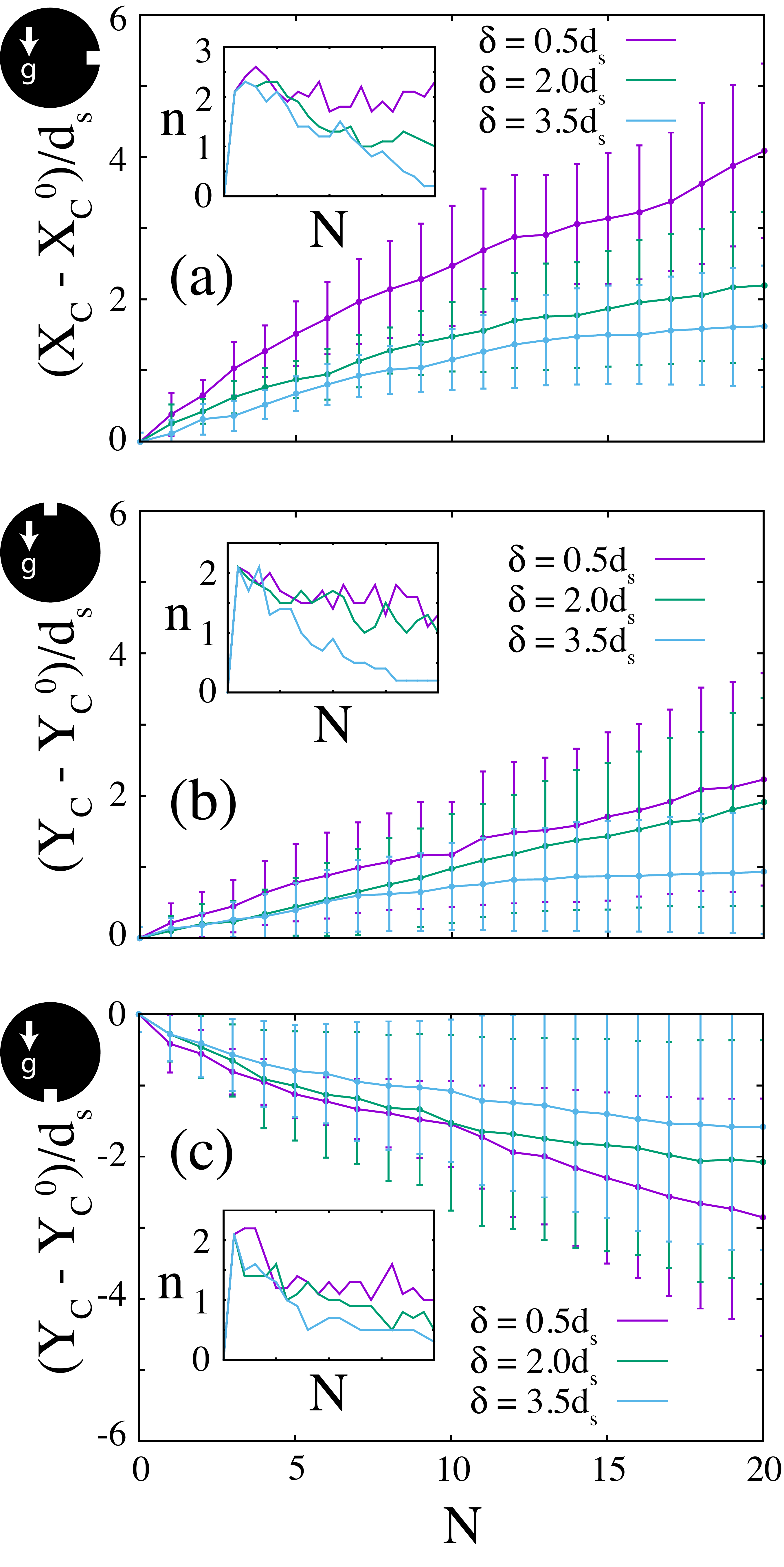}
\caption{\label{fig:mode_all_move} (Color online) The normalized net
  displacement $(X_C-X_C^0)/d_s$ or $(Y_C-Y_C^0)/d_s$ for ejection
  mode (a) perpendicular to gravity, (b) against gravity or (c) with
  gravity as a function of ejection times $N=[1,20]$ and shift
  distance $\delta=0.5d_s$ (dark), $2.0d_s$ (medium) and $3.5d_s$
  (light), respectively. The particle-ejection strength is inversely
  proportional to $\delta$. The insets show the corresponding number
  of ejected particles $n$ with the same horizontal axis. Each curve
  is obtained by averaging the results of ten different initial
  configurations.}
\end{figure}

To evaluate the cruising capability when the cruiser successively
ejects particles entering its square indent-region, we prepare ten
different initial configurations for each particle-ejection mode and
take the average of $(X_C-X_C^0)/d_s$ or $(Y_C-Y_C^0)/d_s$ for the
horizontal or vertical ejection mode as a function of ejection times
$N=[1,20]$ and shift distance $\delta=(0.5d_s, 2.0d_s, 3.5d_s)$. We
also record the average number of particles $n$ being ejected at each
$N$. The results are shown in Fig. \ref{fig:mode_all_move}.

As can be seen clearly in Fig. \ref{fig:mode_all_move}(a), without the
interference from gravity, the three curves of $(X_C-X_C^0)/d_s$ are
well separated and varying systematically, where the curve of
$\delta=0.5d_s$ with the strongest ejection strength on top, and the
curve of $\delta=3.5d_s$ with the weakest ejection strength at the
bottom. On the other hand, as the ejection mode turns vertical,
gravity comes into play by decelerating the cruiser during ejection
against it, or accelerating the cruiser during ejection with it. The
gravity effect complicates the kinetic response, and the mere effect
of changing the ejection strength $\delta$ becomes less obvious. As a
result, the three curves of $(Y_C-Y_C^0)/d_s$ are closer to one
another, as shown in Fig. \ref{fig:mode_all_move}(b) and (c). The
error bars in all three cases are somewhat large for two possible
reasons: 1) The ejection directions of the shifted particles could be
scattered, because we do not strictly control it and let the
particle-cruiser interaction force
$\mathord{\buildrel{\lower3pt\hbox{$\scriptscriptstyle\rightharpoonup$}}
  \over f} _{ij}^n$, defined in Eqn.(\ref{object_force}), take care of
it automatically. 2) The system only has about $4000$ particles and
the boundary effect from the container walls cannot be ignored
easily. We plan to use a much larger system size in our next study to
eliminate these artificial effects and to reduce the error
bars. Lastly, as shown in the insets of Fig. \ref{fig:mode_all_move},
the systematic increasing of the average number of particles $n$ being
ejected at each $N$, as the particle-ejection strength $\delta$
decreases from $3.5d_s$ to $0.5d_s$, agrees with the trend of
$(X_C-X_C^0)/d_s$ and $(Y_C-Y_C^0)/d_s$, where in general the cruiser
can take in more pocketed particles if it ejects more forcefully.

%%%%%%%%%%%%%%%%%%%%%%
%%%%%%%%%%%%%%%%%%%%%%

\subsection{Future work and related experimental setup}
\label{future_work}
For future work, we plan to introduce interparticle friction into our
MD simulation, and explore the stress distribution around the cruiser
to extract the optimal geometry of the indent-region that offers the
best propelling efficiency.

One great advantage of the simple design of the cruiser and the
simulation setup is we can also verify our numerical results with
corresponding experiments. The shifting of pocketed particles in the
indent-region can be done either manually by hand or automatically by
a machine. We can perform the particle-ejection using a mechanical
device that expels the shifted particles using compressed air or a
spring, or impact from a falling object transferring its gravitational
potential energy to the kinetic energy of the ejected particles and
the cruiser.

\section{Conclusions}
\label{conclusions}
In this study, we propose an extremely simple design of a fully-mobile
cruiser in a 2D granular medium under gravity. The cruiser has an
overall round shape except with a square indentation on its edge,
measuring two small particles wide and placed facing the designated
motion direction. In addition to this simple shape, the cruiser can
transfer the particles pocketed in the indent-region to its rear-half
and eject them to obtain thrust for moving forward. The emptied
indent-region offers necessary space to locally unjam the particles
ahead of the cruiser and make room for it. We orient the indentation
to one side of the cruiser to perform the horizontal particle-ejection
mode perpendicular to gravity. For vertical particle-ejection modes
against or with gravity, we put the indentation on the top of or at
the bottom of the cruiser.

Using MD simulations to study the kinetic response of the cruiser in a
sea of nearly 50-50 bidisperse granular particles interacting via the
purely repulsive linear spring force and velocity-dependent damping
force, we identify three distinct phases during an particle-ejection:
1) Phase I starts from the beginning of the ejection until ejected
particles completely leaving the cruiser. In this phase, the cruiser
speeds up and reaches a maximum velocity while its displacement only
increases slightly. 2) Phase II ends until the cruiser exhausts all of
its momentum built up previously. In this phase, the cruiser keeps
moving due to the inertial effects and obtains the longest
displacement among all three phases. The cruiser eventually is stopped
by the compressed particles in front of it. 3) Phase III is when the
compressed particles start to push back until the system becomes fully
relaxed again. At the end of this phase, the decompressed particles
push the cruiser backwards but on average never back to its original
position. We find the three phases exist universally in all three
particle-ejection modes. Besides, we also verify the cruiser indeed
can move smoothly perpendicular or parallel to gravity by successively
ejecting pocketed particles, using the 'one steps forward, half step
back' strategy containing the above three phases. The cruising
efficiency increases proportionally to the ejection strength and works
at its best if without the interference from gravity.

In summary, we believe the simplicity and the efficiency of our
proposed cruiser offers important physics insights for an object
moving within granular materials, and establish a solid base for
engineering mobile objects in athermal media with practical
applications in the visible future.

\section{acknowledgments}
GJG gratefully acknowledges financial support from Shizuoka University
startup funding.

\bibliography{paper_granular_cruiser}

\end{document}